\documentclass[fleqn, floatfix,nofootinbib,apsrev4-1,pra,reprint,groupedaddress]{revtex4-1}
\pdfoutput=1
\usepackage{natbib}
\usepackage{caption}
\usepackage{amsmath,graphicx}
\usepackage{epstopdf}
\usepackage[english]{babel}
\usepackage[T1]{fontenc}
\usepackage{babel}
\usepackage{float}
\usepackage{amsmath}
\usepackage{graphicx}
\usepackage{subfig}
\usepackage{varioref}
\usepackage[latin9]{inputenc}
\usepackage[unicode=true,pdfusetitle,
 bookmarks=true,bookmarksnumbered=false,bookmarksopen=false,
 breaklinks=false,pdfborder={0 0 1},backref=false,colorlinks=false]
 {hyperref}
 
\makeatletter

\@ifundefined{showcaptionsetup}{}{%
 \PassOptionsToPackage{caption=false}{subfig}}
\usepackage{subfig}
\makeatother

\begin{document}
\title{Criterion for Stability of a Special Relativistically Covariant Dynamical System}
\author{L.P. Horwitz$^{1,2,3}$ and D. Zucker$^1$
\\
${}^1$ $School\:of\:Physics\:and\: Astronomy, Raymond\: and \:Beverly \:Sackler \:Faculty \:of \:Exact \:Sciences, Tel \:Aviv -University, Ramat \:Aviv, 69978, Israel$.\\
${}^2$ $Department\: of \:Physics, Bar\: Ilan\: University, Ramat - Gan, 52900, Israel$.\\
${}^3$ $Department\: of \:Physics, Ariel \:university \:in \:Samaria, Ariel, 44837 \:Israel$.}

{\let\thefootnote\relax\footnotetext{email addresses: larry@post.tau.ac.il (L.P. Horwitz)\\ zucker@mail.tau.ac.il (D. zucker)}}
 
\begin{abstract}
We study classically the problem of two relativistic particles with an invariant Duffing-like potential which reduces to the usual Duffing form in the nonrelativistic limit. We use a special relativistic generalization (RGEM) of the geometric method (GEM) developed for the analysis of nonrelativistic Hamiltonian systems to study the local stability of a relativistic Duffing oscillator. Poincar\'e plots of the simulated motion are consistent with the RGEM. We find a threshold for the external driving force required for chaotic behavior in the Minkowski spacetime. 
\end{abstract}

\maketitle

\section{Introduction}

In Newton's classical mechanics and in quantum mechanics, one makes
use of a global time that has causal meaning. The manifestly covariant
Stueckelberg formalism \cite{Stueckelberg:1941rg},\cite{17.horwitz1973relativistic} is based on
a similar idea, that there is an invariant parameter $\tau$ of evolution
of the system. 

According to Stueckelberg, covariant functions of space and time (Einstein's
time, $t$) form a Hilbert space (over $R^4$) for each value of $\tau$
(which we shall refer to as ``world time''). Thus, there are two
types of time, one transforming covariantly, and the second an invariant parameter
of evolution\cite{horwitz2015relativistic}. The first type, Einstein's time in the theory of relativity,
$t$, of an event as measured in the laboratory is subject to variation (Lorentz transformation)
according to the velocity of the apparatus related to the transmitting
system and may as well be affected by forces (such as gravity), and
the second type is the Stueckelberg's world time, $\tau$, that remains
unaffected \cite{horwitz1988two}.

This allows us to write equations of motion for a single particle
that depend on the world time  $\tau$, in terms of Hamilton equations based
on a Hamilton function of $x,p$ in the quantum theory where $x,p$ are covariant four-vectors.
The dynamics of the system is driven by $\tau$; $t$ and $x$ are quantities that are observed in the laboratory.  The Maxwell equations, for examples, describe electromagnetic phenomena in terms of the observable $t$ and $x$, but. in Stueckelberg's framework, the sources (with spacetime positions given by $x$ and $t$) are governed by equations of motion in $\tau$, as discussed in detail in \cite{17.horwitz1973relativistic}. The results of our analysis therefore describe the physically observable consequences of the dynamical evolution. consequences of the dynamical evolution.

As done by Stueckelberg, one can write a Schr$\ddot{o}$dinger type equation
for the wave function for a free particle (for $\mu = 0,1,2,3$):

\[
i\frac{\partial}{\partial\tau}\psi_{\tau}(x^{\mu})=\eta_{\mu\nu}\frac{p^{\mu}p^{\nu}}{2M}\psi_{\tau}(x^{\mu}),
\]
where $x^{\mu}=x(x,t)$, $\eta_{\mu\nu}= diag (-1,1,1,1)$ and $\tau$ is the invariant parameter of evolution.
M is a parameter of dimension mass (but not the measured or the rest mass;
it may be considered the Galilean (nonrelativistic) limit mass \cite{11.horwitz1981gibbs}).
One sees that for the classical case by the Hamilton equations

\[
\dot{x}^{\mu}\equiv\frac{ dx^{\mu}}{d\tau}=\frac{p^{\mu}}{M}=\frac{\partial H}{\partial p_{\mu}}
\]
so that

\[
-\frac{ds^{2}}{d\tau^{2}}=\frac{dx^{\mu}dx_{\mu}}{d\tau^{2}}=-\frac{m^{2}}{M^{2}}
\]
where $m$ is the measured mass and $s$ is the proper time. Therefore, the proper time coincides with with $\tau$ when $m=M$ (called the "mass shell").  

To describe the dynamical evolution of a more than one particle system,
Horwitz and Piron generalized Stueckelberg's theory and postulated
a universal $\tau$ for any number of particles, enabling the solution
of the two body problem in particular. They separated variables to
CM (center of four-mometnum frame variables) and relative motion, and solved the classical Kepler problem with invariant potential proportional to $1/\rho$, where $\rho=\sqrt{x^{2}-t^{2}}$,
with $x,t$ relative coordinates. Arshansky and Horwitz \cite{12.arshansky1989quantum1},
reduced the relativistic quantum problem to the corresponding nonrelativistic Schr$\ddot{o}$dinger equation with $V(\rho)$, where $\rho$ replaces the radial coordinate r everywhere in the equation.
The solutions they found were irreducible representations of O(2,1)
and they therefore introduced an induced representation to obtain
representations of O(3,1) \cite{13.arshansky1989quantum2}. 

Horwitz and Schieve treated the reduced motion of a relativistic
Duffing oscillator with a potential of the form 

\[
V(\rho)=a\rho^{4}-b\rho^{2}+c
\]

where $\rho^{2}=x^{\mu}x_{\mu}$, by simulation. 

In this work, we generalize the method of Horwitz et al \cite{14.horwitz2007geometry} (called GEM) for studying the stability\footnote{Here we are discussing local stability in the neighborhood of points on the orbit, often associated with the overall stability of the system. We use the term "stability" throughout this paper in this sense. except for our discussion of Poincare plots, which reflect the overall stability of the system.} of nonrelativistic Hamiltonian systems to the relativistic
case (which we shall call RGEM), and use both methods, simulation and the application of the
RGEM criterion to study the stability of the relativistic Duffing oscillator.

It will be shown here that the unstable orbits found by Schieve and Horwitz pass through the regions of instability indicated by the GEM criterion generalized to the relativistic (RGEM) case and that the occurrence of chaos on Poincar\'e plots and on the space time orbits.
The formula for the relativistic generalization is derived here.
This application is the first that has been carried out for the relativistic form of this criterion.

\section{Review of Nonrelativistic Stability Theory}

It has been shown \cite{13.arshansky1989quantum2} that a Hamiltonian of standard form in nonrelativistic
mechanics $(for i,j,k = 1,2,3)$:

\begin{equation}\label{eq:Hamiltonian_flat_space}
H=\frac{1}{2M}\eta_{\text{ij}}p^{i}p^{j}+V(y)
\end{equation}

can be cast \cite{14.horwitz2007geometry} into a geometrical form
\cite{15.gutzwiller2013chaos} (a method which we call GEM):

\begin{equation}\label{eq:Hamiltonian_curved_space}
\hat{H}=\frac{1}{2M}g_{\text{ij}}p^{i}p^{j}
\end{equation}

where $g_{ij}=\phi(x)\delta_{ij}$ (we shall call the variable of
the curved space manifold $x_{i}$ to distinguish from the Hamilton
space designated here by$y^{i}$).

The Hamilton equations applied to (\ref{eq:Hamiltonian_curved_space}) imply the geodesic equation (a local diffeomorphism covariant form)

\begin{equation}
\ddot{x}_{i}=-\Gamma_{i}^{jk}\dot{x}_{j}\dot{x}_{k}
\end{equation}

where 

\begin{equation}
\Gamma_{i}^{jk}=\frac{1}{2}g_{im}(\frac{\partial g^{mk}}{\partial x_{j}}+\frac{\partial g^{jm}}{\partial x_{k}}-\frac{\partial g^{jk}}{\partial x_{m}})
\end{equation}

A mapping in the tangent space defined by

\begin{equation}
\dot{x}_{i}=g_{ij}(x)\dot{y}^{j}
\end{equation}

puts this equation into the (also) diffeomorphism covariant form

\begin{equation}\label{eq:connetion_form}
\ddot{y}^{i}=-\frac{1}{2}g^{ij}\frac{\partial g_{kl}}{\partial y^{j}}\dot{y}^{k}\dot{y}^{l}\equiv-M_{kl}^{i}\dot{y}^{k}\dot{y}^{l}
\end{equation}

In the special coordinate system (for $H=\hat{H}=E$) for which $g_{ij}=\phi\delta_{ij}$,$\phi=E/(E-V)$,
this equation reduces to: 
\begin{equation}
\ddot{y}^{j}=-\frac{\partial V(y)}{\partial y^{j}},
\end{equation}
 and therefore recovers the usual Hamiltonian evolution.

We shall identify the variables $y^{i}$with the Hamiltonian variable,
since they constitute in (\ref{eq:connetion_form}) an embedding of the actual Hamilton
motion into a curved space.

As a measure of stability, the geodesic deviation derived from this is given by

\begin{equation}
\frac{D^{2}\zeta^{i}}{D\tau^{2}}=R_{jkl}^{i}\dot{y}^{\text{j}}\dot{y}^{l}\zeta^{k},
\end{equation}

where $R_{jkl}^{i}$ is a curvature associated with the geodesic motion, turns out to be a very sensitive and reliable criterion for the stability
of the original system\cite{16.zion2008applications}. In this work,
we generalize this procedure to the relativistic case.

\section{The relativistic Hamiltonian and its transformation to geometric form}

Let us consider a Hamiltonian in Minkowski space of the form:

\begin{equation}\label{flat_space_Hamiltonian}
H=\eta_{\mu\nu}\frac{p^{\mu}p^{\nu}}{2M}+V(y)
\end{equation}

where $\eta_{\mu\nu}$ is the Minkowski metric and the manifold is
described in the $y$-coordinate system by our definition. 

A Hamiltonian of this kind is assumed by Stueckelberg \cite{stueckelberg1993helv},
who formulated the basic dynamics for both classical and quantum theory. Horwitz and Piron \cite{17.horwitz1973relativistic} generalized the theory to be applicable to the many body problem, by defining the parameter 
$\tau$ as universal, and worked out the two body Kepler case as an example. The quantum two body problem
with invariant action at a distance potential was solved by Horwitz
and Arshansky \cite{18.arshansky1988solutions}.

Our goal is to check the stability of the system by applying a conformal
transformation on the metric and writing the Hamiltonian in a curved
space (as done for the nonrelativistic case in \cite{14.horwitz2007geometry};
see also Gershon and Horwitz \cite{19.gershon2009kaluza}).

To do so, we will first write an \textquotedbl{}equivalent\textquotedbl{}
Hamiltonian in curved space with no potential \cite{20.misner1973gravitation}: 

\begin{equation}
\hat{H}=g_{\mu\nu}\frac{p^{\mu}p^{\nu}}{2M}
\end{equation}

where $g_{\mu\nu}$ is the metric of the new manifold which we will
describe in a coordinate manifold which we shall call ${x}$; to establish
the equivalence, we require $p^{\mu}(\tau)$ to be the same in both
systems. We define this dynamical equivalence on the basis that the measureable quantities $p^\mu$, associated with forces, should be the same at every stage of the evolution governed by $\tau$; it enables the definition of the conformal map \cite{14.horwitz2007geometry} between $(1)$ and $(2)$, shown to be highly effective in the nonrelativistic case.

In order to construct a relation between $H$ and $\hat{H}$, let us define a conformal transformation on the metric, i.e. (on the manifold ${x}$)

\begin{equation}
g_{\mu\nu}=\phi(x)\eta_{\mu\nu}
\end{equation}

We shall assume the ``energy'' $E$ (value of $H$) to be equal for both
systems and a correspondence between the $x$'s and $y$'s so that\footnote{The units of $H$ and $\hat{H}$ here are that of mass, but we shall use the
familiar appellation of energy. In fact the energy of the system is the time component of the {\it total} four-momentum, $p_{1}^{\mu}+p_{2}^{\mu}$.}:

\begin{equation}
E=\eta_{\mu\nu}\frac{p^{\mu}p^{\nu}}{2m}+V(y)=g_{\mu\nu}(x)\frac{p^{\mu}p^{\nu}}{2M}
\end{equation}

consequently:
\begin{equation}\label{eq:phi}
\phi(x)=\frac{E}{E-V(y)}
\end{equation}

for the corresponding points $x$ and $y$.\footnote{L.P. Horwitz, A. Yahalom, J. Levitan and M. Lewkowitch [to be published] have shown that with (13) and (15), all derivatives of (14) can be expressed in terms of derivatives of $\phi(x)$, and conversely, so the two manifolds are
well defined in an analytic domain.}

It follows from the requirement of equal momentum that (the dot corresponds
to differentiation with respect to $\tau$): 

\begin{eqnarray}\label{eq:defnition}
\dot{y}^{\mu} & = & g^{\mu\nu}(x)\dot{x}_{\nu}\\
\dot{x}_{\nu} & = & g_{\mu\nu}(x)\dot{y}^{\mu}
\end{eqnarray}

so that: 

\begin{equation}
\ddot{y}^{\mu}=\frac{\partial g^{\mu\nu}}{\partial\tau}\dot{x}_{\nu}+g^{\mu\nu}\ddot{x}_{\nu}\\
\end{equation}

and also, by the Hamilton equations derived from (9) \cite{15.gutzwiller2013chaos},\cite{14.horwitz2007geometry}

\begin{equation}
\ddot{x}_{\alpha}  =  -\frac{1}{2}g_{\alpha\nu}(\frac{\partial g^{\nu\theta}}{\partial x_{\lambda}}+\frac{\partial g^{\theta\nu}}{\partial x_{\lambda}}-\frac{\partial g^{\lambda\theta}}{\partial x_{\nu}})\dot{x}_{\lambda}\dot{x}_{\theta} = - \Gamma_\alpha^{\lambda \theta} \dot x_\lambda {\dot x}_\theta    
\end{equation}

so that: 
\begin{equation}\label{eq:ddot_y}
\ddot{y}_{\nu}=\frac{1}{2}\frac{\partial g^{\lambda\theta}}{\partial x_{\nu}}g_{\rho\lambda}g_{\sigma\theta}\dot{y}^{\rho}\dot{y}^{\sigma}\equiv-M_{\rho\sigma}^{\mu}\dot{y}^{\rho}\dot{y}^{\sigma}
\end{equation}

We remark that since ${y}$ corresponds to a manifold different from ${x}$ (through the mapping (\ref{eq:defnition})), we have called the corresponding connection form $M_{\rho \sigma}^\mu$.

Now let us define the deviation between two orbits by
\begin{eqnarray}
\zeta{}^{\alpha} & = & y'^{\alpha}-y^{\alpha}\\
y'^{\alpha} & = & y^{\alpha}+\zeta^{\alpha}
\end{eqnarray}

so that

\begin{equation}
\ddot{y}'^{\nu}=-M^{'}{}_{\rho\sigma}^{\nu}\dot{y}'^{\rho}\dot{y}'^{\sigma}
\end{equation}

where:

\begin{equation}
M^{'}{}_{\rho\sigma}^{\nu}=M_{\rho\sigma}^{\nu}+\partial_{\alpha}(M_{\rho\sigma}^{\nu})\zeta^{\alpha}
\end{equation}

and:

\begin{equation}
\partial_{\alpha}=\frac{\partial}{\partial y^\alpha}
\end{equation}

Through the definition (\ref{eq:phi}),$M_{\rho\sigma}^{\nu}$ is explicitly a function of $y^{\alpha}$.

Now, expanding up to first order in $\zeta$, $\dot{\zeta}$ we get:

\begin{equation}\label{eq:ddot_zeta}
\ddot{\zeta}^{\nu}=-2M{}_{\rho\sigma}^{\nu}\dot{y}{}^{\rho}\dot{\zeta}^{\sigma}-\partial_{\alpha}(M{}_{\rho\sigma}^{\nu})\dot{y}{}^{\rho}\dot{y}{}^{\sigma}\zeta^{\alpha}
\end{equation}

If $\zeta$ is small enough, we can treat it as a tensor, and its
covariant derivative is: 

\begin{equation}
\frac{D\zeta^{\nu}}{D\tau}=\dot{\zeta}^{\nu}+M{}_{\rho\sigma}^{\nu}\dot{y}^{\rho}\zeta^{\sigma}
\end{equation}

The second covariant derivative is 

\begin{equation}\label{eq:cov_derivative}
\frac{D^{2}\zeta^{\nu}}{D\tau^{2}}=\frac{d}{d\tau}(\dot{\zeta}^{\nu}+M{}_{\rho\sigma}^{\nu}\dot{y}^{\rho}\zeta^{\sigma})+M{}_{\alpha\beta}^{\nu}(\dot{\zeta}^{\alpha}+M{}_{\rho\sigma}^{\alpha}\dot{y}^{\rho}\zeta^{\sigma})\dot{y}^{\beta}\\
\end{equation}

Substituting (\ref{eq:ddot_zeta}) into (\ref{eq:cov_derivative}), we get

\begin{eqnarray}\label{eq:D_cov_derivative}
\frac{D^{2}\zeta^{\nu}}{D\tau^{2}}&=&\nonumber-\partial_{\alpha}(M{}_{\rho\sigma}^{\nu})\dot{y}{}^{\rho}\dot{y}{}^{\sigma}\zeta^{\alpha}+\partial_{\alpha}(M{}_{\rho\sigma}^{\nu})\dot{y}^{\alpha}\dot{y}^{\rho}\zeta^{\sigma}\\
&+&M{}_{\rho\sigma}^{\nu}\ddot{y}^{\rho}\zeta^{\sigma}+M{}_{\alpha\beta}^{\nu}M{}_{\rho\sigma}^{\alpha}\dot{y}^{\beta}\dot{y}^{\rho}\zeta^{\sigma}
\end{eqnarray}

Now inserting (\ref{eq:ddot_y}) into (\ref{eq:D_cov_derivative}) and changing contracted indices we get:

\begin{equation}
\frac{D^{2}\zeta^{\nu}}{D\tau^{2}}=R_{\beta\sigma\rho}^{\nu}\dot{y}^{\beta}\dot{y}^{\rho}\zeta^{\sigma}
\end{equation}

where $R$ is a \textquotedbl{}dynamical curvature\textquotedbl{} associated with the geodesic motion (\ref{eq:ddot_y}), i.e;
\begin{equation}
R_{\mu\nu\sigma}^{\alpha}=\frac{\partial M_{\mu\nu}^{\alpha}}{\partial y^{\sigma}}-\frac{\partial M_{\mu\sigma}^{\alpha}}{\partial y^{\nu}}+M_{\mu\nu}^{\lambda}M_{\sigma\lambda}^{\alpha}-M_{\mu\sigma}^{\lambda}M_{\nu\lambda}^{\alpha}
\end{equation}

In the special coordinate system in which (\ref{eq:phi}) is valid
\begin{eqnarray}
R_{\mu\sigma\nu}^{\alpha}=\nonumber\eta^{\alpha\kappa}[\frac{1}{2\phi(y)}(\eta_{\nu\mu}\frac{\partial^{2}\phi(y)}{\partial y^{\sigma}\partial y^{\kappa}}-\eta_{\sigma\mu}\frac{\partial^{2}\phi(y)}{\partial y^{\nu}\partial y^{\kappa}})&&\\
-\frac{1}{4\phi^{2}(y)}(\eta_{\mu\nu}\frac{\partial\phi(y)}{\partial y^{\sigma}}\frac{\partial\phi(y)}{\partial y^{\kappa}}-\eta_{\sigma\mu}\frac{\partial\phi(y)}{\partial y^{\nu}}\frac{\partial\phi(y)}{\partial y^{\kappa}})]&&
\end{eqnarray}

Using (\ref{eq:phi}), we can calculate the first and second derivatives of $\phi(y)$ and consequently we can calculate R explicitly:

\begin{equation}\label{eq:curvature_tensor}
\begin{aligned}
&R_{\mu\nu\sigma}^{\alpha}=& \\ 
&\eta^{\alpha\kappa}[\frac{3}{4(E-V(y))^{2}}(\eta_{\mu\nu}\frac{\partial V(y)}{\partial y^{\sigma}}\frac{\partial V(y)}{\partial y^{\kappa}}-\eta_{\sigma\mu}\frac{\partial V(y)}{\partial y^{\nu}}\frac{\partial V(y)}{\partial y^{\kappa}})&\\
&+\frac{1}{2(E-V(y))}(\eta_{\mu\nu}\frac{\partial^{2}V(y)}{\partial y^{\sigma}\partial y^{\kappa}}-\eta_{\sigma\mu}\frac{\partial^{2}V(y)}{\partial y^{\nu}\partial y^{\kappa}})]&
\end{aligned}
\end{equation}

from (\ref{flat_space_Hamiltonian}), we have
\begin{eqnarray}\label{eq:Hamiltonian}
E-V(y) & = & \eta_{\mu\nu}\frac{p^{\nu}p^{\nu}}{2m}=\frac{p^{2}}{2M}
\end{eqnarray}

We also know from the Hamilton equations that

\begin{equation}\label{eq:EOM}
\dot{y}^{\mu} = \frac{\partial H}{\partial p_{\mu}}= \frac{1}{M}p^{\mu}
\end{equation}

Substituting (\ref{eq:Hamiltonian}) and (\ref{eq:EOM}) into (\ref{eq:curvature_tensor}),and after contracting $\mu$ and $\sigma$, the geodesic deviation equation then reads: 

\begin{eqnarray}
&&\eta_{\alpha\kappa}\frac{D^{2}\zeta^{\alpha}}{D\tau^{2}}=\nonumber \\
&&[\eta_{\mu\nu}(\frac{3}{M^{2}\dot{y}^{2}}\frac{\partial V(y)}{\partial y^{\sigma}}\frac{\partial V(y)}{\partial y^{\kappa}}+\frac{1}{M}\frac{\partial^{2}V(y)}{\partial y^{\sigma}\partial y^{\kappa}})\frac{\dot{y}^{\mu}\dot{y}^{\sigma}}{\dot{y}^{2}}\nonumber \\
&&-(\frac{3}{M^{2}\dot{y}^{2}}\frac{\partial V(y)}{\partial y^{\nu}}\frac{\partial V(y)}{\partial y^{\kappa}}+\frac{1}{M}\frac{\partial^{2}V(y)}{\partial y^{\nu}\partial y^{\kappa}})]\zeta^{\nu}
\end{eqnarray}

Let us define 

\begin{equation}\label{matrix}
Q_{\sigma\kappa}=\frac{3}{M^{2}\dot{y}^{2}}\frac{\partial V(y)}{\partial y^{\sigma}}\frac{\partial V(y)}{\partial y^{\kappa}}+\frac{1}{M}\frac{\partial^{2}V(y)}{\partial y^{\sigma}\partial y^{\kappa}}
\end{equation}

and the projection (orthogonal to ${\dot p}^\mu$)
\begin{equation}
p_{\nu}^{\sigma}=(\eta_{\nu}^{\sigma}-\eta_{\mu\nu}\frac{\dot{y}^{\mu}\dot{y}^{\sigma}}{\dot{y}^{2}}).
\end{equation}

we can then write
\begin{eqnarray}\
\eta_{\alpha\kappa}\frac{D^{2}\zeta^{\alpha}}{D\tau^{2}} & = & -p_{\nu}^{\sigma}Q_{\sigma\kappa}\zeta^{\nu}
\end{eqnarray}

We wish to project the deviation on the direction orthogonal to ${\dot y}^\mu$. We then obtain

\begin{eqnarray}
p_{\alpha}^{\gamma}\frac{D^{2}\zeta^{\alpha}}{D\tau^{2}} & = & -(p_{\alpha}^{\gamma}Q_{\sigma}^{\alpha}p_{\nu}^{\sigma})\zeta^{\nu}
\end{eqnarray}

If the eigenvalues of $Q_{\sigma}^{\alpha}$ are
positive, then the eigenvalues of $p_{\alpha}^{\gamma}Q_{\sigma}^{\alpha}p_{\nu}^{\sigma}$
are positive as well, so that if both of the eigenvalues of $Q_{\sigma}^{\alpha}$ are positive, then the system is stable.
If either or both of the eigenvalues are negative, then the system may be considered unstable. In this case, there could be an exponentially rapid divergence of neighboring trajectories, charfacteristic of local instability. A zero eigenvalue may be considered as indicating instability as well, since a small additional perturbation may render it negative.

\section{The Duffing Oscillator}

$ $

We now choose a potential of the form of the invariant relativistic
Duffing oscillator \cite{23.schieve1991chaos}:
\begin{equation}\label{Duffing}
\centering
V(y)=ay^{4}-by^{2}+c,
\end{equation}

where $y^\mu = y_1^\mu -y_2^\mu$, the relative coordinates of the
two body system and we take $c=0$ in this study.

We shall study this problem for the 1+1 dimensional case. As we have
seen, stability depends on the eigenvalues of the resulting matrix
obtained by projecting the left hand side also in the direction orthogonal
to the trajectory.  
Inserting (\ref{Duffing}) into (\ref{matrix}) will result in the following symmetric matrix of the form: 

\begin{equation}
Q=\left[\begin{array}{cc}
A & B\\
B & C
\end{array}\right]
\end{equation}

where
\begin{eqnarray}
A&=&-(\frac{3}{M^{2}\dot{y}^{2}}(4ay^{2}-2b)^{2}+\frac{1}{M}8a)y^{0}y_{0}\nonumber \\
&&+\frac{1}{M}(4ay^{2}-2b)\\
B&=&(\frac{3}{M^{2}\dot{y}^{2}}(4ay^{2}-2b)^{2}+\frac{1}{M}8a)y_{1}y^{0}\\
C&=&(\frac{3}{M^{2}\dot{y}^{2}}(4ay^{2}-2b)^{2}+\frac{1}{M}8a)y^{1}y_{1}\nonumber \\
&&+\frac{1}{M}(4ay^{2}-2b)
\end{eqnarray}

In the general case of a symmetric matrix of this form the eigenvalues are: 

\begin{equation}
\lambda_{1,2}=\frac{(A+C)\pm\sqrt{(A+C)^{2}+4(B^{2}-AC)}}{2}
\end{equation}

Since the eigenvalues of a real symmetric matrix are real, we obtain two conditions which the matrix elements must meet in order for both eigenvalues to be positive:  

1. $(A+C)>0$

2. $4(B^{2}-AC)<0$\\

If the one of the conditions fails then at least one of the eigenvalues
is negative. 

The first condition gives is:
\begin{equation}
(\frac{3}{M^{2}\dot{y}^{2}}(4ay^{2}-2b)^{2}+\frac{1}{M}8a)y^{2}+\frac{2}{M}(4ay^{2}-2b)>0
\end{equation}

Using (\ref{eq:Hamiltonian}) and (\ref{eq:EOM}) we can write the first condition explicitly:

\begin{equation}
(\frac{3}{8(E-V(y))}(4ay^{2}-2b)^{2}+4a)y^{2}>b
\end{equation}

The second condition gives: 
\begin{equation}\label{second_condition}
\begin{aligned}
\frac{1}{M}(4ay^{2}-2b)(\frac{3}{M^{2}\dot{y}^{2}}(4ay^{2}&-2b)^{2}+\frac{1}{M}8a)y^{2}>&\\
&-\frac{1}{M^{2}}(4ay^{2}-2b)^{2}&
\end{aligned}
\end{equation}

where we used the relation $y_{1}y^{0}=-y_{0}y^{1}$. 
Using the fact that $\frac{1}{M^{2}}(4ay^{2}-2b)^{2}>0$, and substituting (\ref{eq:Hamiltonian}) and (\ref{eq:EOM}) into (\ref{second_condition}) ,we can write the second condition explicitly: 

\begin{equation}
(\frac{3(4ay^{2}-2b)}{2(E-V(y))}+\frac{8a}{(4ay^{2}-2b)})y^{2}+1>0
\end{equation}

Next, the equations of motion with respect to $\tau$ for the Hamiltonian of the form

\begin{equation}\label{eq:Duffing_Hamiltonian}
E\equiv H=\frac{p_{\mu}p^{\mu}}{2M}+ay^{4}-by^{2}
\end{equation}

are
\begin{equation}\label{eq:ydot}
\ddot{y}^{\mu}=\frac{\dot{p}^{\mu}}{M}=\frac{(4ay^{2}-2b)\eta^{\alpha\mu}y_{\alpha}}{M}
\end{equation}

\section{Results}

We wrote an algorithm to draw the regions of instability and to solve the equations of motion numerically.
the algorithm was a variation of the MidPoint method, only that in our case we didn't possess the first derivative of $y$, we had to evaluate it in every iteration along with the value of $y$ itself. 
The regions of instability depend only on the value of $E$ and are
independent of the initial conditions. 
The smallest eigenvalue occurs where both conditions are met.
Setting $E=1$ and adding a driving force the equations of motion now read \footnote{In $1+1$ dimensions, $H$ and the Lorentz generators are conserved, so there is no chaotic motion without a driving force.} 
\begin{equation}
\ddot{y}^{\mu}=\frac{\dot{p}^{\mu}}{M}=\frac{(4ay^{2}-2b)\eta^{\alpha\mu}y_{\mu}}{M}+f\text{y}^{\mu}sin(\omega\tau)
\end{equation}
Explicit equations for the time and space parts of $y$ are obtained by selecting the indices $0$ and $i$ for $\mu$.
We examine the results that follow from the conditions for stability and the trajectories that correspond to different values of the driving force coefficient.
Also, using the period of $\omega$ we can construct a Poincare
plot for $\rho$ and $\dot{\rho}$, where $\rho=\sqrt{x^{2}-t^{2}}$ and $\dot{\rho}=\frac{x\dot{x}-t\dot{t}}{\rho}$.
We wish to study whether chaotic behavior can be detected and what are the conditions for generating it. 
Setting  \footnote{These constants are chosen to correspond to the values taken in \cite{23.schieve1991chaos}, but the results remain qualitatively the same for other choices.} $a=\frac{3}{4}$ , $b=\frac{1}{2}$ and the initial conditions $t=0.001$ , $x=0.2$ , $\dot{t}=0.001$ and $\dot{x}=0.01$ to comply with the paper of Horwitz and Schieve \cite{23.schieve1991chaos},we have 3 parameters to choose arbitrarily for the equations of motion: $M,f$ and $\omega$.
We set $\omega=\frac{\pi}{3}$ for convenience. 
Inserting (\ref{eq:ydot}) into (\ref{eq:Duffing_Hamiltonian}) we get: 
\begin{equation}
E=H=\frac{M\dot{y}^{2}}{2}+ay^{4}-by^{2}+c
\end{equation}

For $E=1$ and no driving force, i.e, $f=0$, we obtain the results in (fig.1).

\begin{figure}
\subfloat[]{\includegraphics[width=0.8\columnwidth, height=0.8\columnwidth]{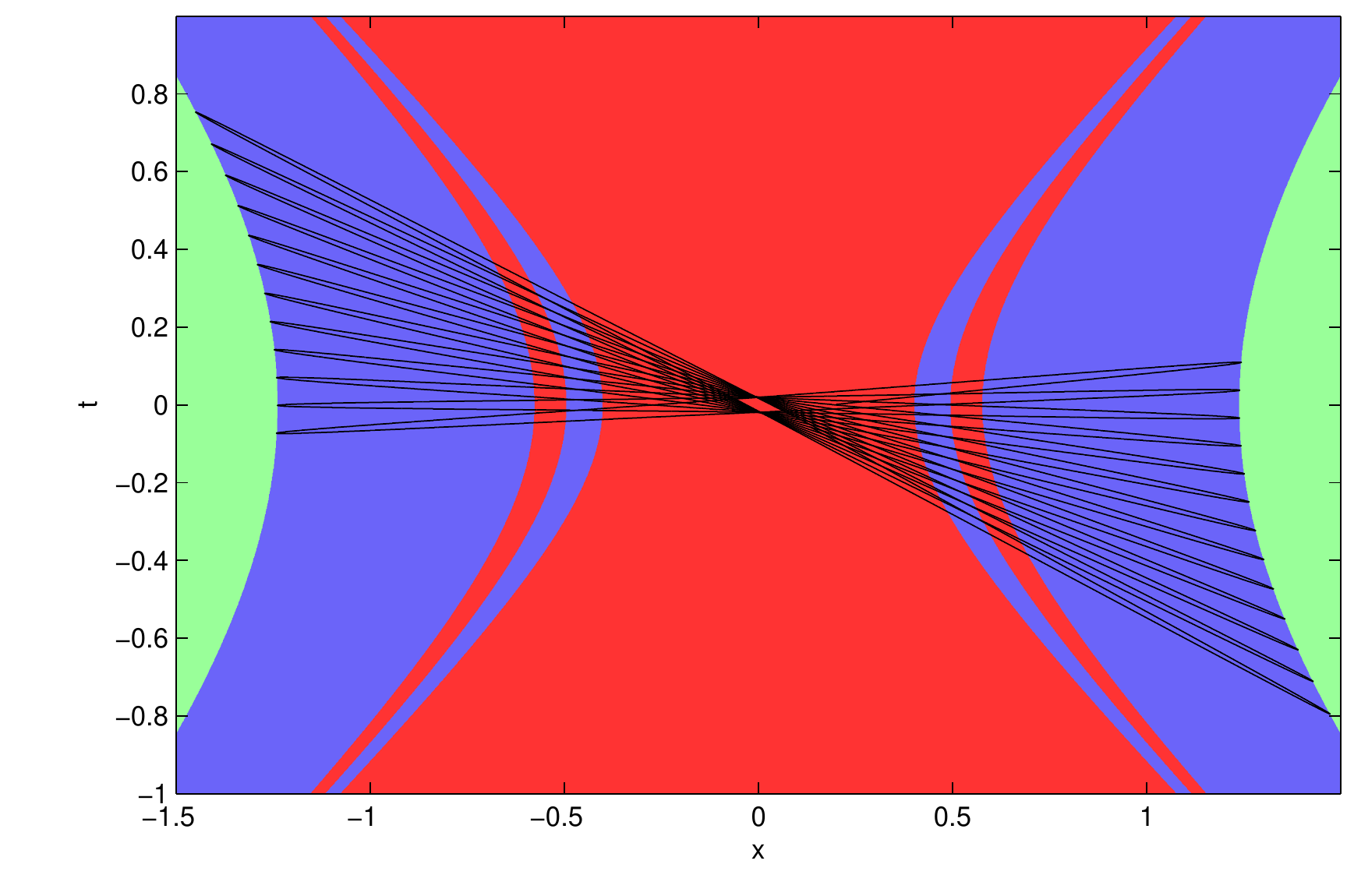}

}\hfill{}\subfloat[]{\includegraphics[width=0.8\columnwidth, height=0.8\columnwidth]{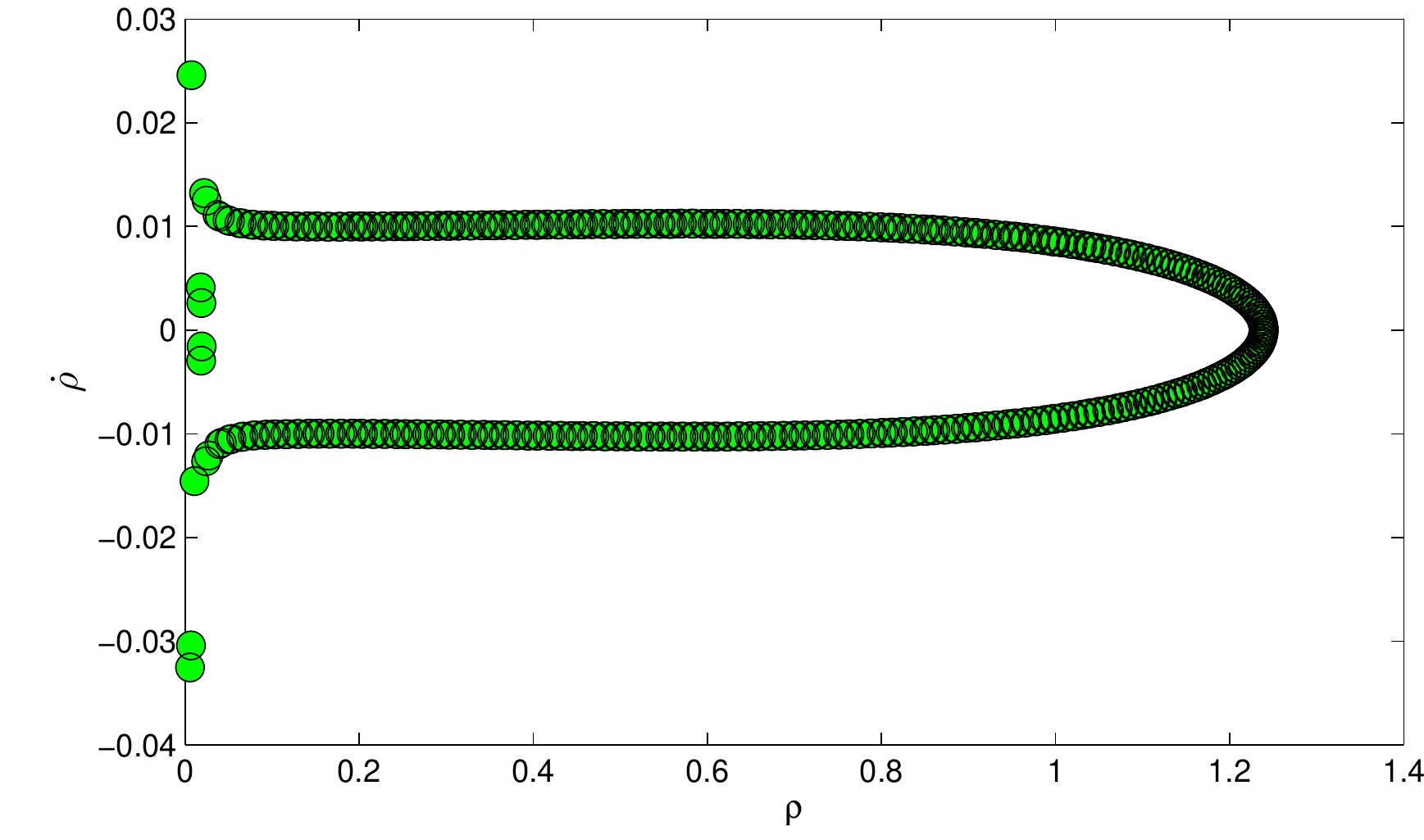}}
\caption{(color online) (a) The trajectory in the x-t plain. The blue area is the stable area. The red area is an unstable area where only one of the conditions for instability is met. 
The green area is an unstable area where both conditions of instability are met and the instability is the strongest. (b) The Poincare plot of the motion using $\omega$ as the period of the motion.}
\end{figure}

We can see that the trajectory crosses the red areas but doesn't go
into the green area where the instability is the strongest.
We can also see that in the blue area, the trajectory doesn't change
its direction or crosses itself, however when the trajectory enters
the unstable region around the origin, trajectory crosses itself;
and when it reaches the green area, it changes its direction. 
This pattern will be conserved until the trajectory runs off to
infinity. \\
We can also see that the Poincar\'e plot doesn't indicate chaotic behavior since the trajectory crosses the $\rho-\dot{\rho}$ plan in a different point every period but there is a distinct pattern to these crossings  and for the case of $f=0$, it is also symmetric around $\dot{\rho}=0$.

Other plots we wish to examine are those of the trajectory in each dimension as a function of $\tau$ (fig.2).
\begin{figure}
\subfloat[]{\includegraphics[width=0.8\columnwidth, height=0.8\columnwidth]{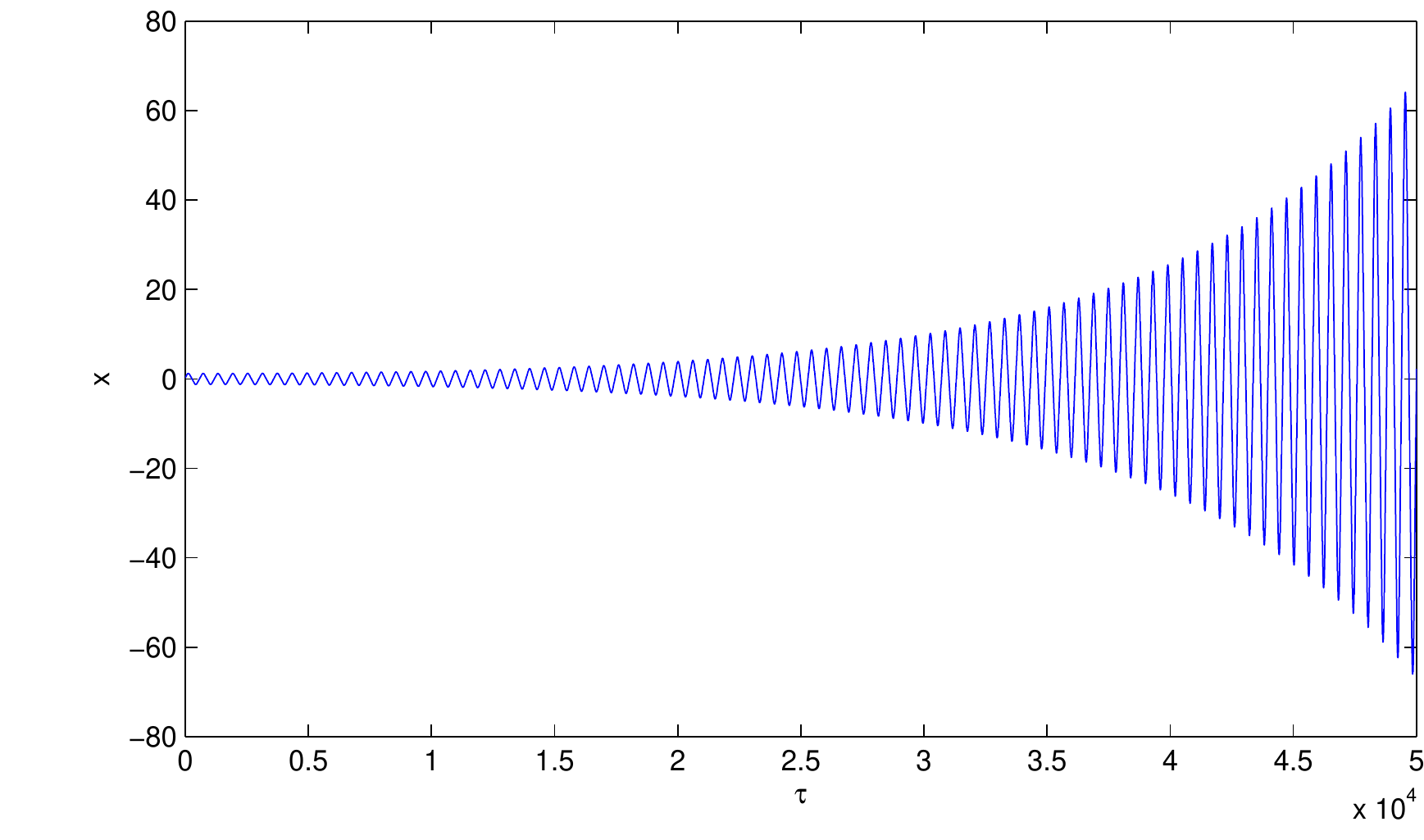}

}\hfill{}\subfloat[]{\includegraphics[width=0.8\columnwidth, height=0.8\columnwidth]{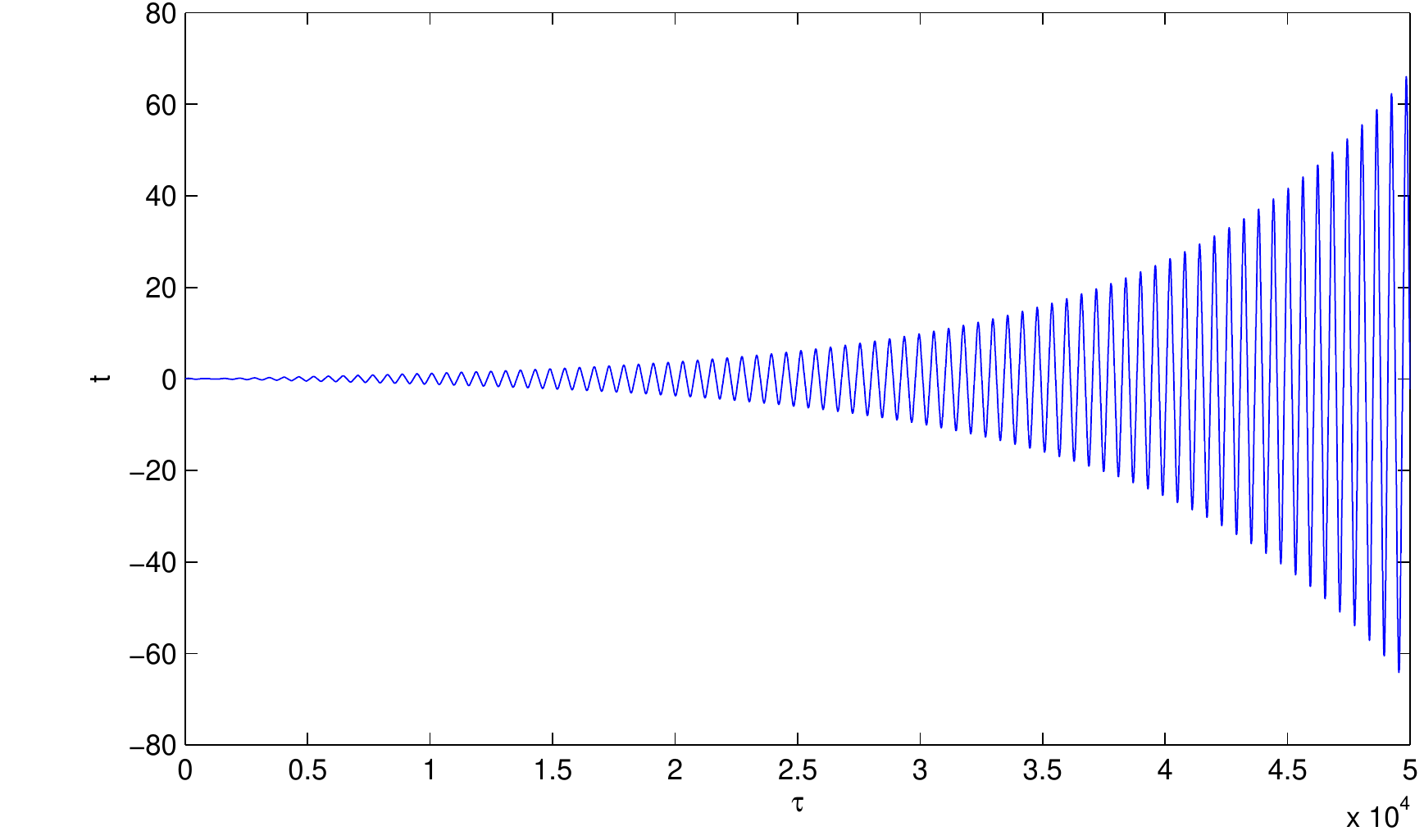}}
\caption{Both x (a) and t (b) as functions of $\tau$ shows expanding oscillations with a distinct pattern and what seems to be a constant period in $\tau$. }
\end{figure}
The trajectories on the x-$\tau$ and t-$\tau$ planes have distinct patterns and they do not exhibit chaotic behavior.  

We now add a driving force and set $f=0.05$ and examine the results (fig.3).

\begin{figure}
\subfloat[]{\includegraphics[width=0.8\columnwidth, height=0.8\columnwidth]{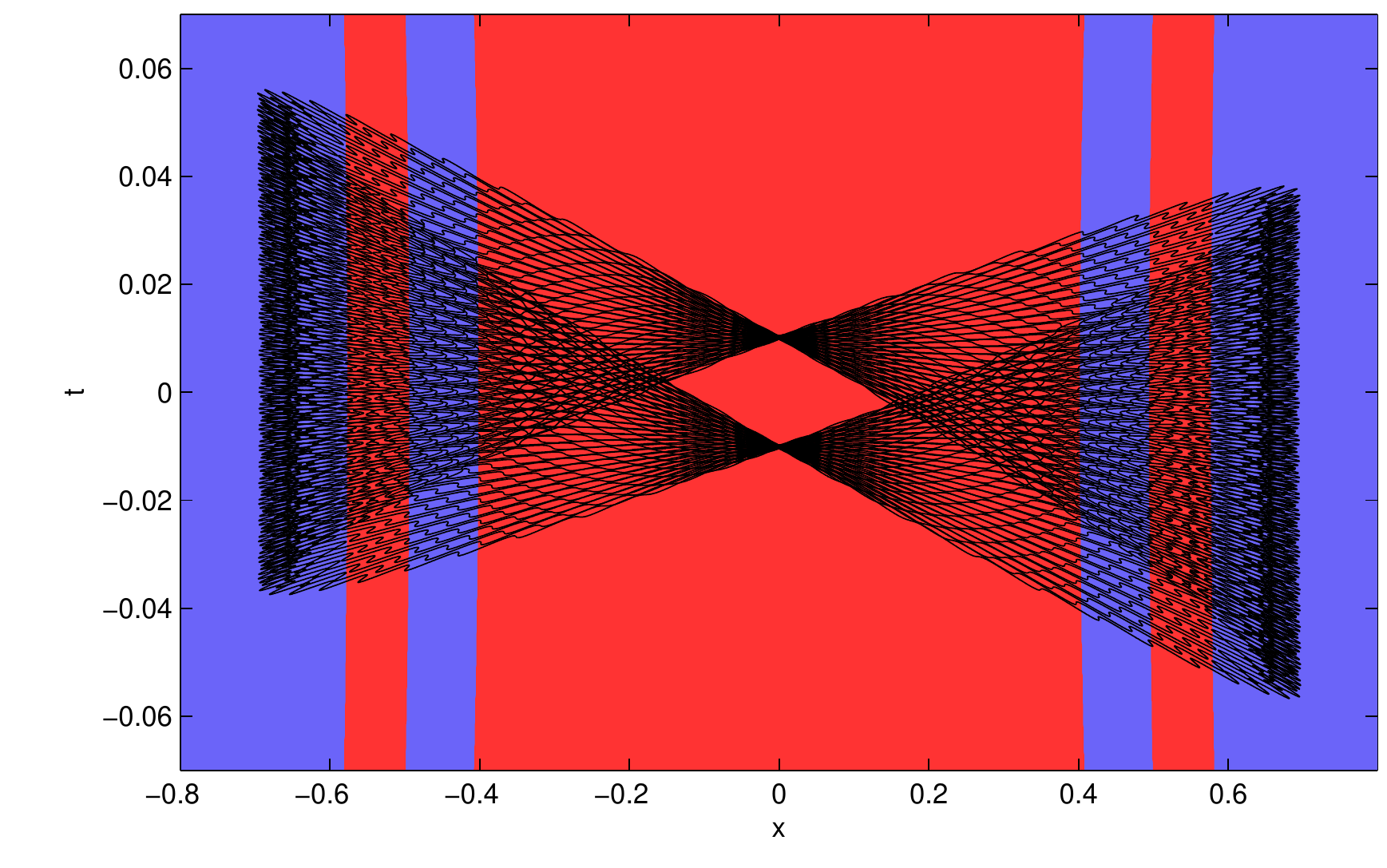}

}\hfill{}\subfloat[]{\includegraphics[width=0.8\columnwidth, height=0.8\columnwidth]{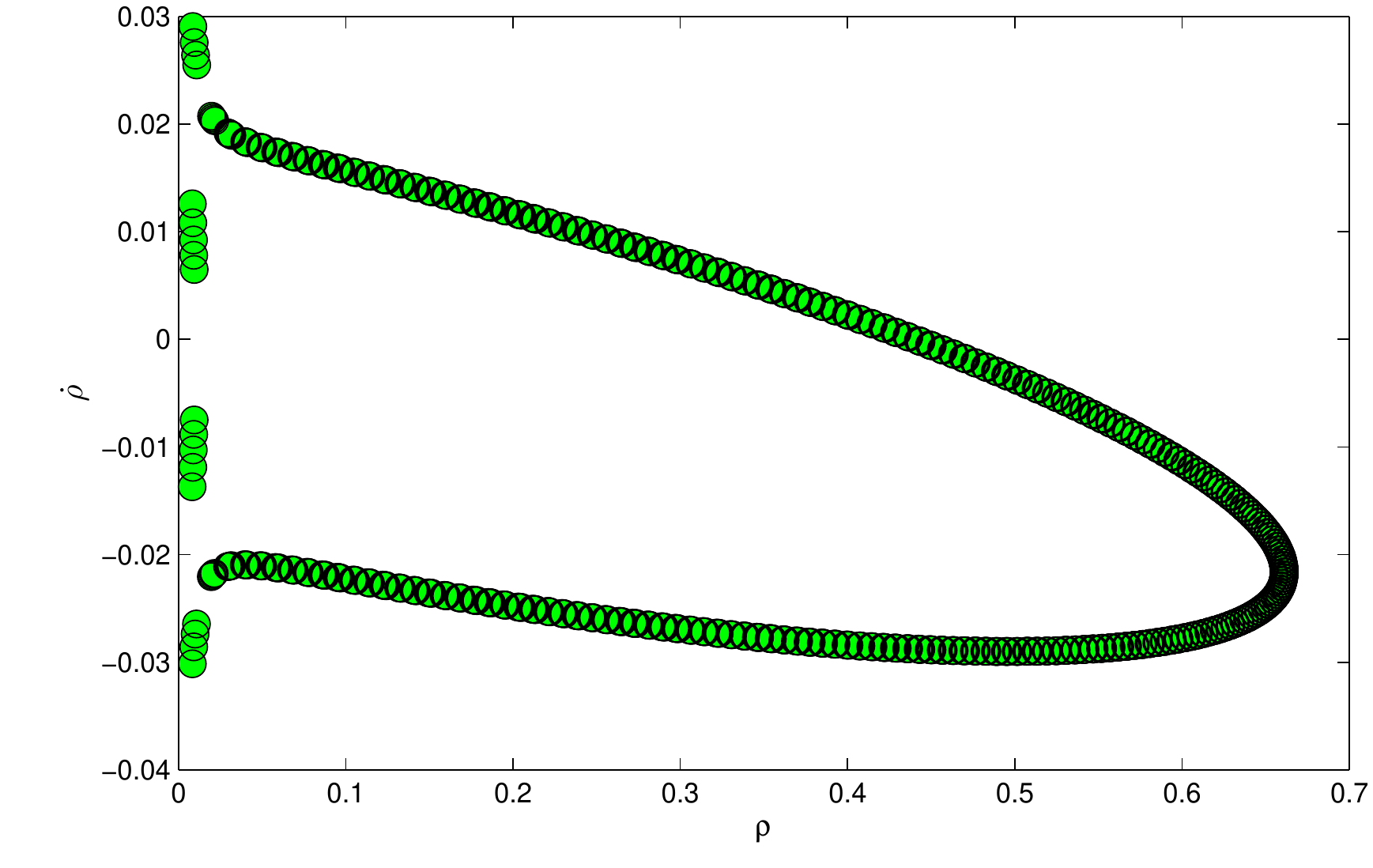}}
\caption{The motion under the influence of a small driving force, is contained within a small area around the origin and crosses the unstable regions there (a).  
We can also see that the trajectory crosses the $\rho-\dot{\rho}$ plan
in a different point every period but there is a distinct pattern
to these crossings (b).}
\end{figure}

We can see that by adding this small driving force, the trajectory
stays around the origin and doesn't spread.
We can also see that the trajectory crosses the unstable red area but doesn't
go into the green area. 
This motion doesn't indicate chaotic behavior.
The fact that at each period, the crossing of the $\rho-\dot{\rho}$
plane occurs on a different point and never twice at the same point
implies unstable behavior, however, there is a distinct pattern spread
over a very small area which does not indicate chaotic behavior, although the plot is no longer symmetric around $\dot{\rho}=0$ due to the driving force. 

Examining the the trajectory in each dimension as a function of $\tau$ we get the results in (fig.4).
\begin{figure}
\subfloat[]{\includegraphics[width=0.8\columnwidth, height=0.8\columnwidth]{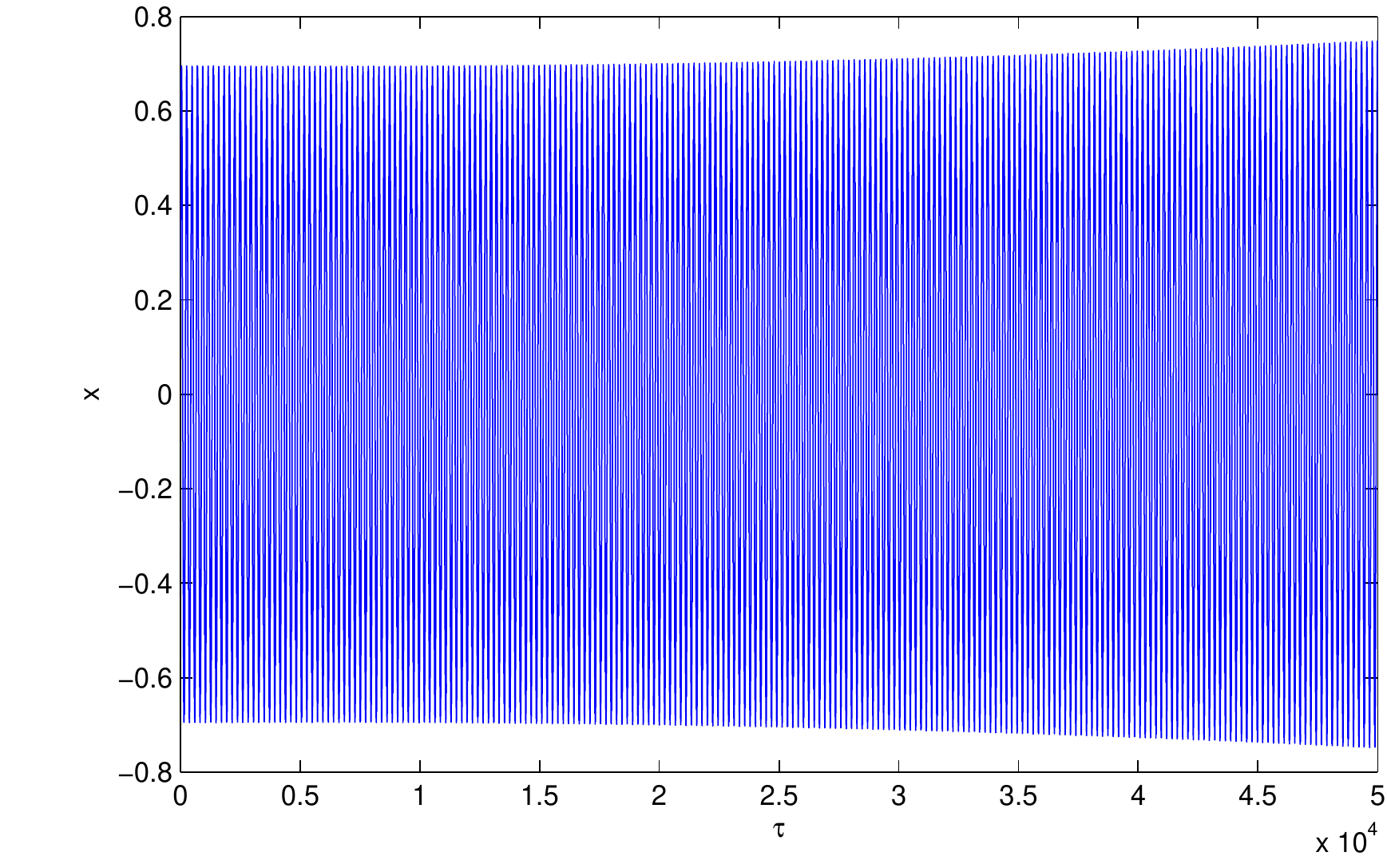}

}\hfill{}\subfloat[]{\includegraphics[width=0.8\columnwidth, height=0.8\columnwidth]{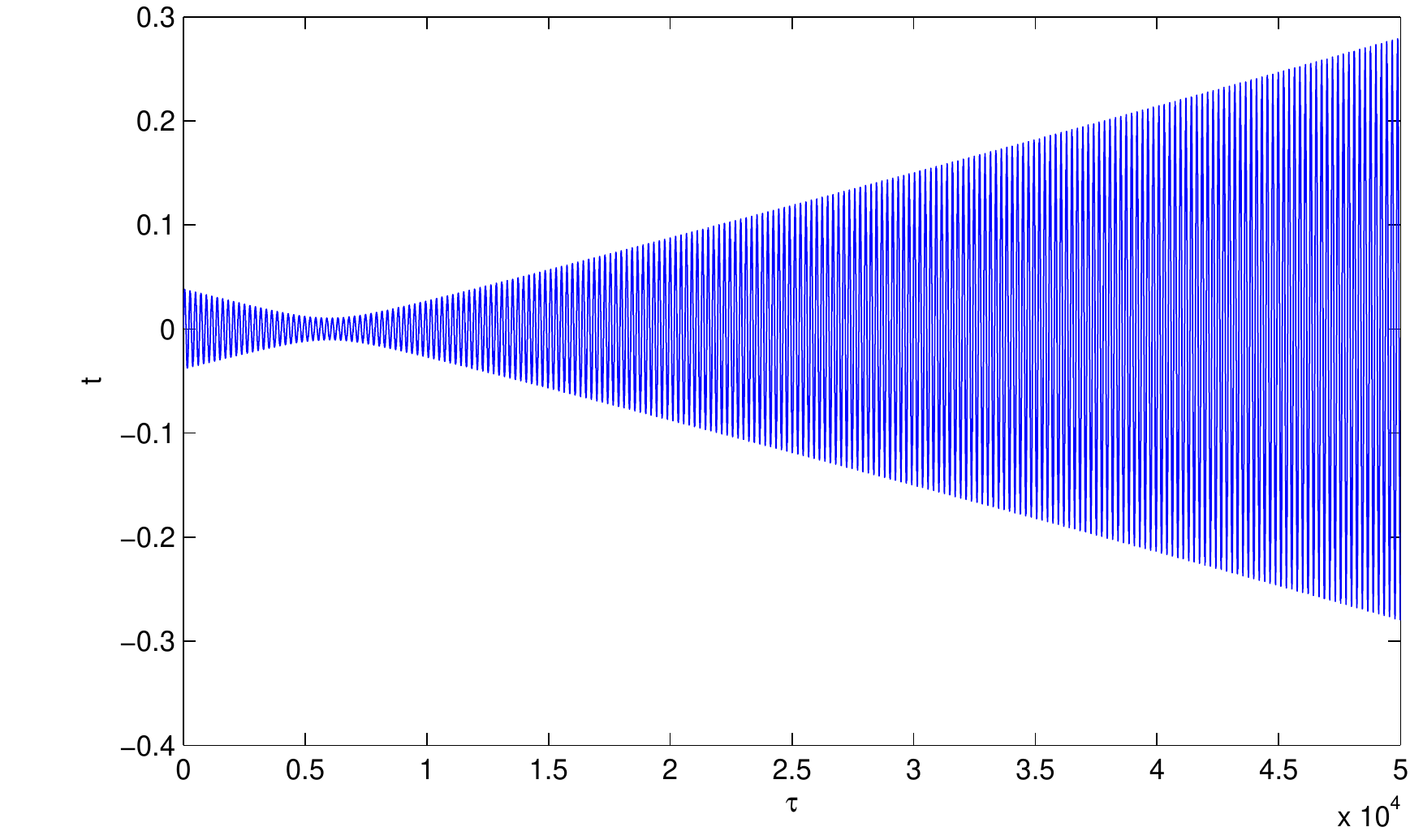}}
\caption{Both x (a) and t (b) as functions of $\tau$ shows oscillations with a distinct pattern and what seems to be a constant period in $\tau$.}
\end{figure}

The trajectories of $t$  as a function  of $\tau$ snd $x$ as a function of $\tau$ do not imply such behavior as well.
We see that there is a distinct pattern of these plots and what seems to be a constant period of the motion. 

If we take f=1 (fig.5), 
\begin{figure}
\subfloat[]{\includegraphics[width=0.8\columnwidth, height=0.8\columnwidth]{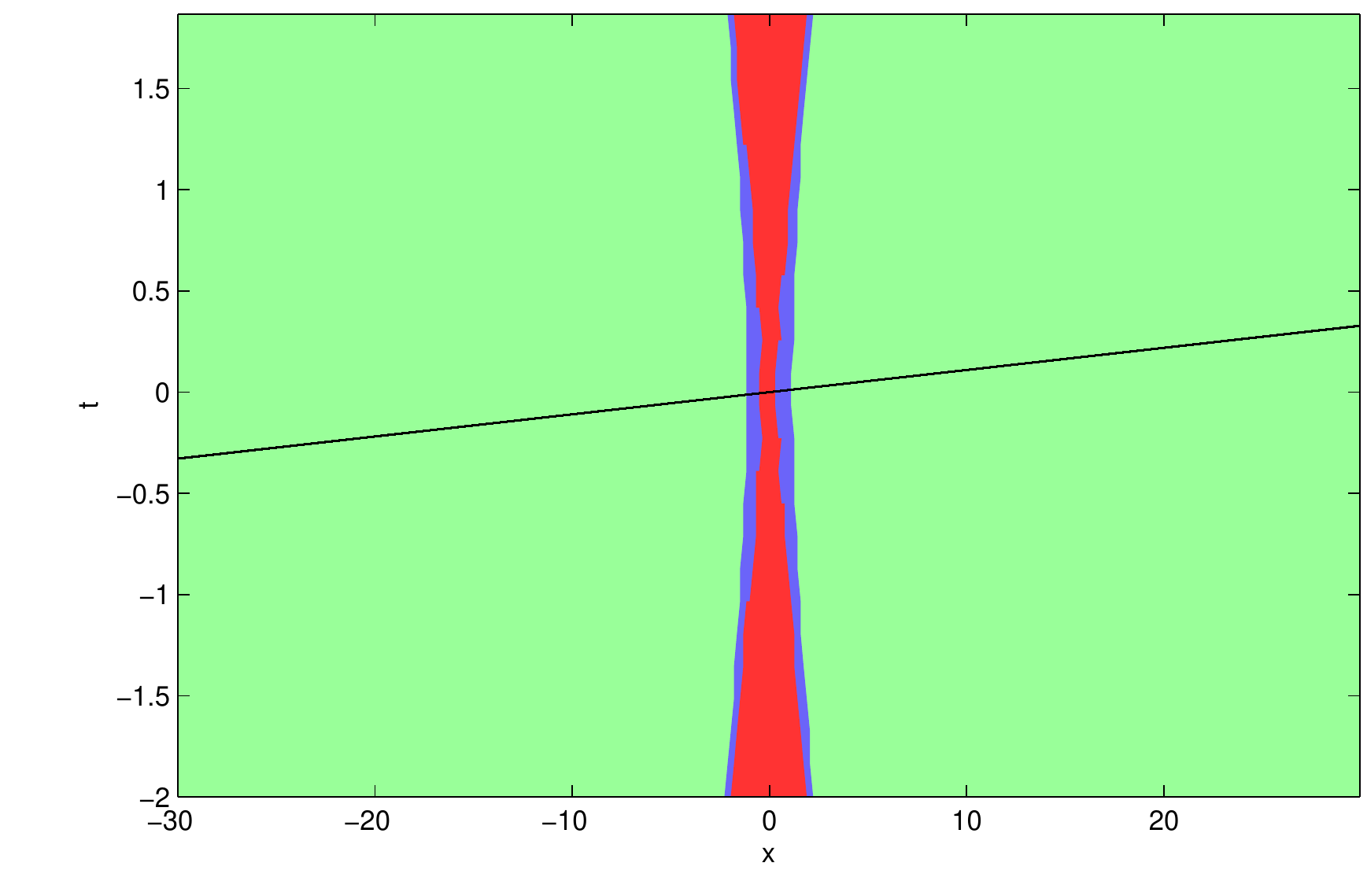}

}\hfill{}\subfloat[]{\includegraphics[width=0.8\columnwidth, height=0.8\columnwidth]{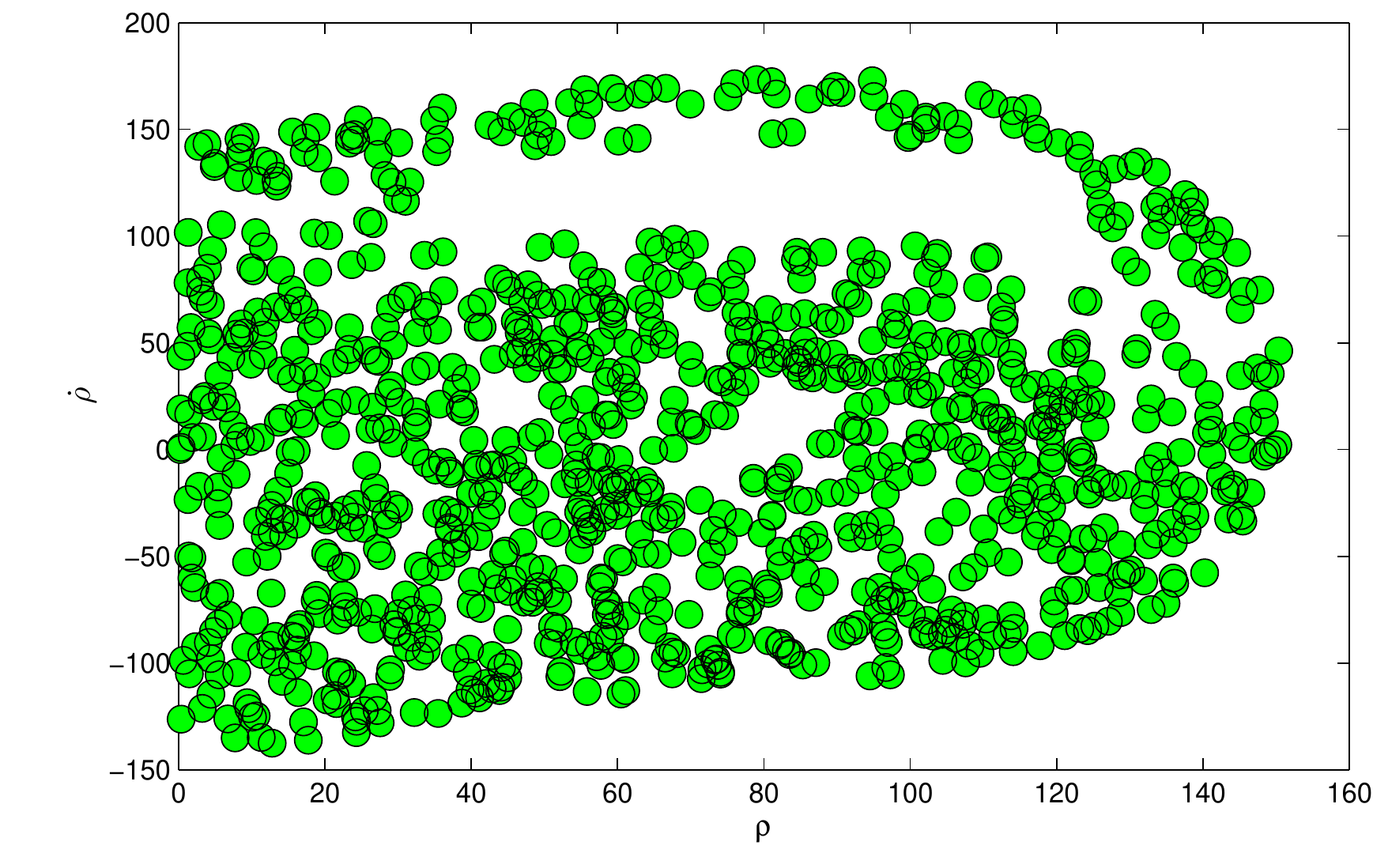}}

\caption{(a)The trajectory we get from the equations of motion for $E=1$ and
$f=1$ over the different regions. (b)The Poincare plot for $E=1$ and $f=1$ using $\omega$ as the period
of the motion. Approximatelty $1000$ points are represented in this figure.}
\end{figure}
we see that the trajectory (fig.5a) oscillates in a wide range of
x and goes deep into the green area where the instability is the strongest. 
The Poincare plot (fig.5b) is scattered randomly with no distinct pattern, suggesting chaotic behavior. We find no regularity in the resulting plots.

If we focus on the trajectory and enlarge it significantly, we will
see that there are 2 sets of trajectories (fig.6).

\begin{figure}
\includegraphics[width=0.8\columnwidth, height=0.8\columnwidth]{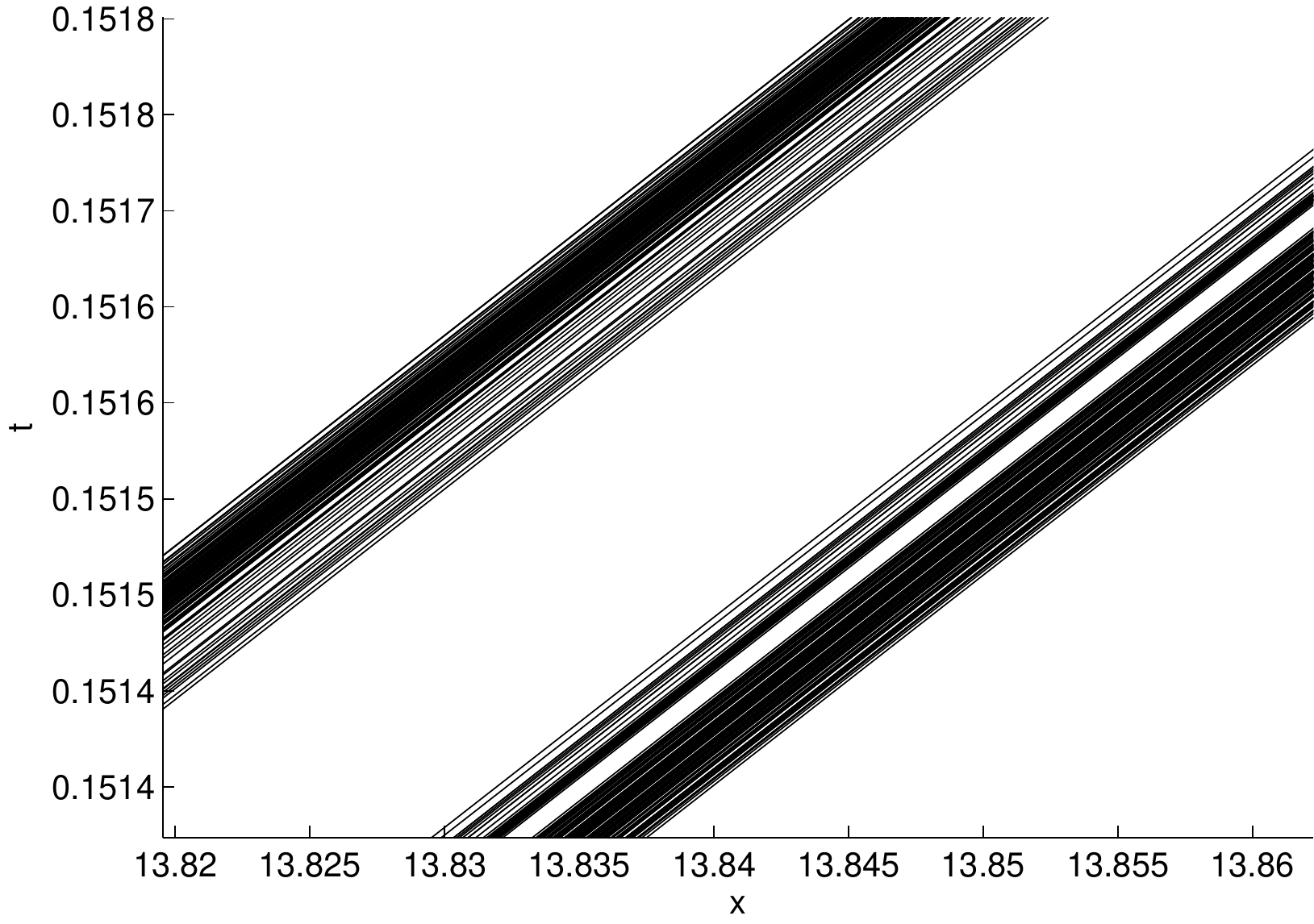}

\caption{A closer look at the trajectory reveals two distinct subsets}
\end{figure}

Each contains a group of close-by sub-trajectories - which matches
the two potential wells of the Duffing oscillator and it looks like
a period doubling which suggests chaotic behavior. 

Also, the trajectories of $x$ vs $\tau$ (fig.7a) and $t$ vs $\tau$
(fig.7b) imply such behavior.Both plots indicate chaotic behavior.

\begin{figure}
\subfloat[]{\includegraphics[width=0.8\columnwidth, height=0.8\columnwidth]{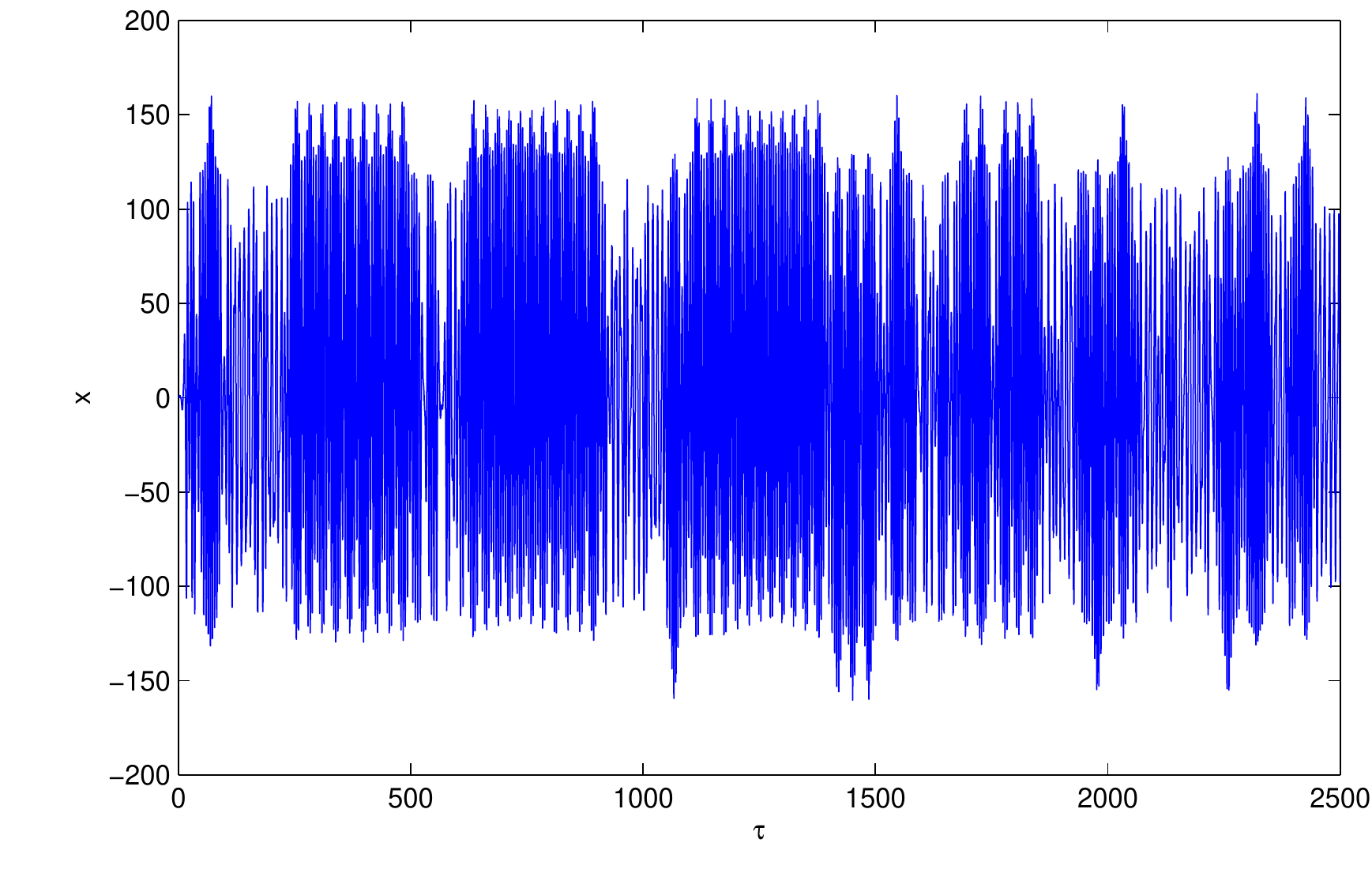}

}\hfill{}\subfloat[]{\includegraphics[width=0.8\columnwidth, height=0.8\columnwidth]{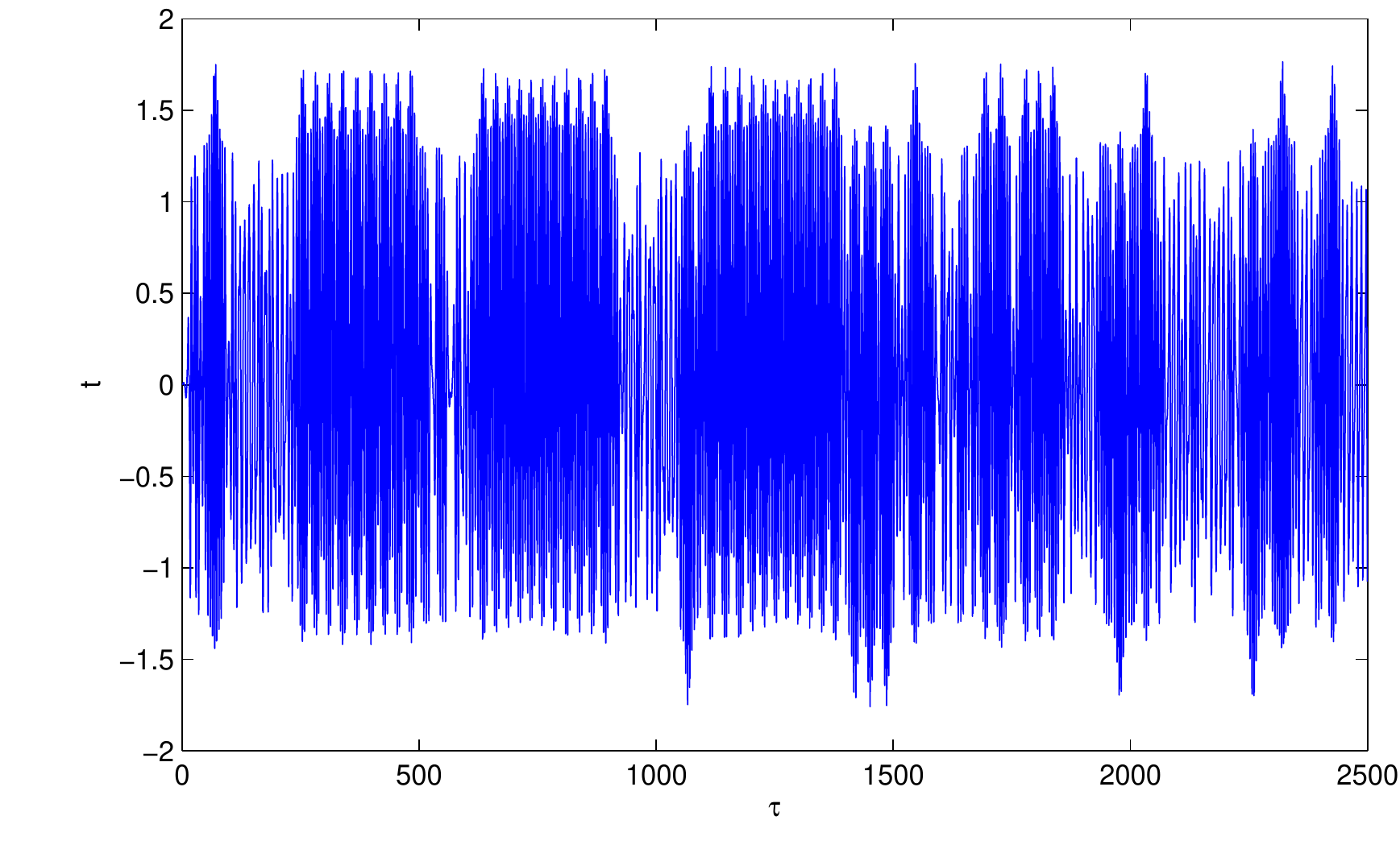}}
\caption{Both x (a) and t (b) as functions of $\tau$ shows no distinct pattern or a period in $\tau$.}
\end{figure} 
The value of the driving force that seems to be the boundary between
stability and instability for E=1 appear to be f=0.71, i.e., larger values provide clearly unstable behavior, and smaller values, clearly stable. For this choice, we obtain (fig.8) and (fig.9), demonstrating what might be expected of a threshold behavior.

In this case, we obtain
(fig.8).

\begin{figure}
\subfloat[]{\includegraphics[width=0.8\columnwidth, height=0.8\columnwidth]{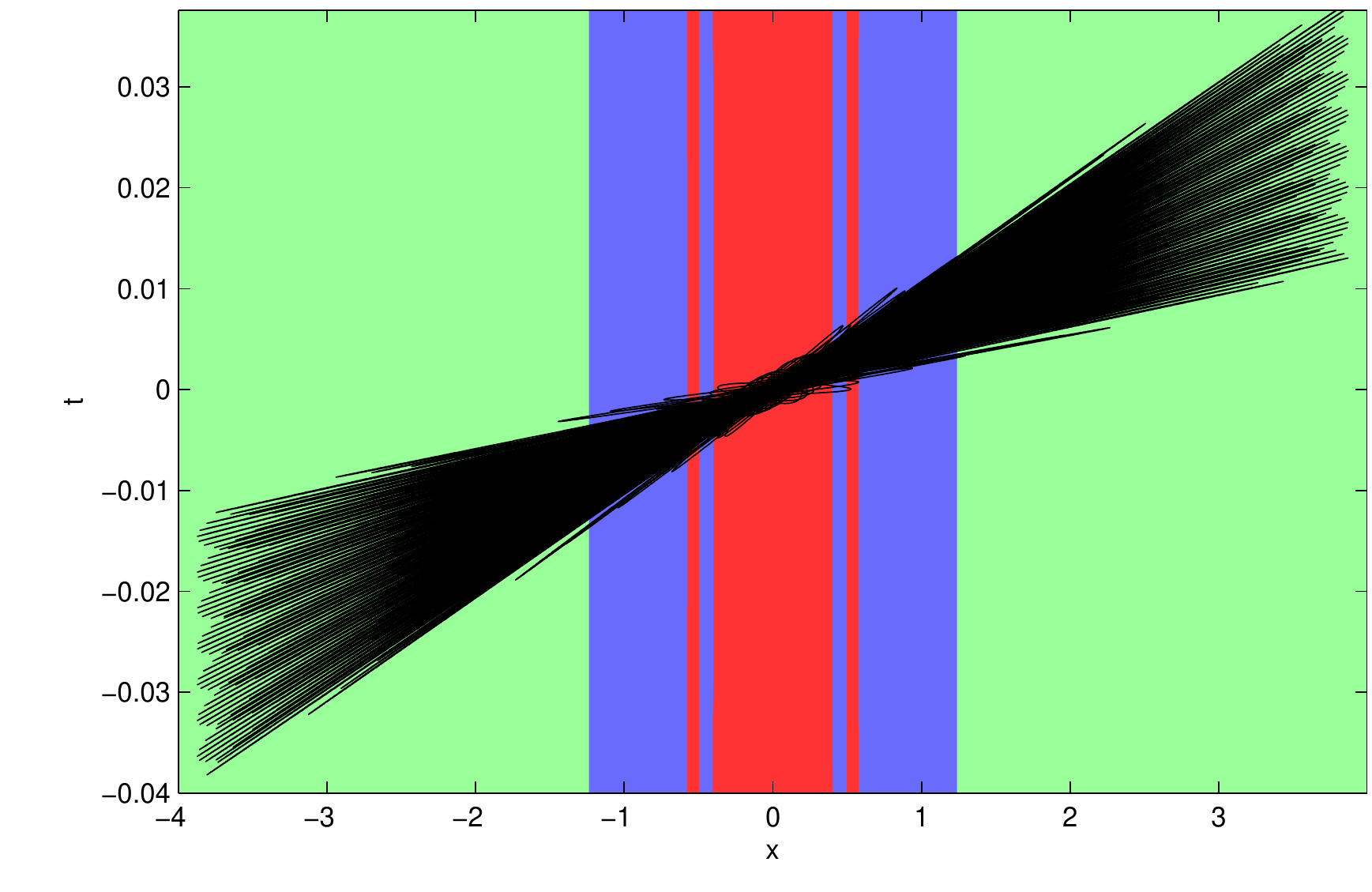}

}\hfill{}\subfloat[]{\includegraphics[width=0.8\columnwidth, height=0.8\columnwidth]{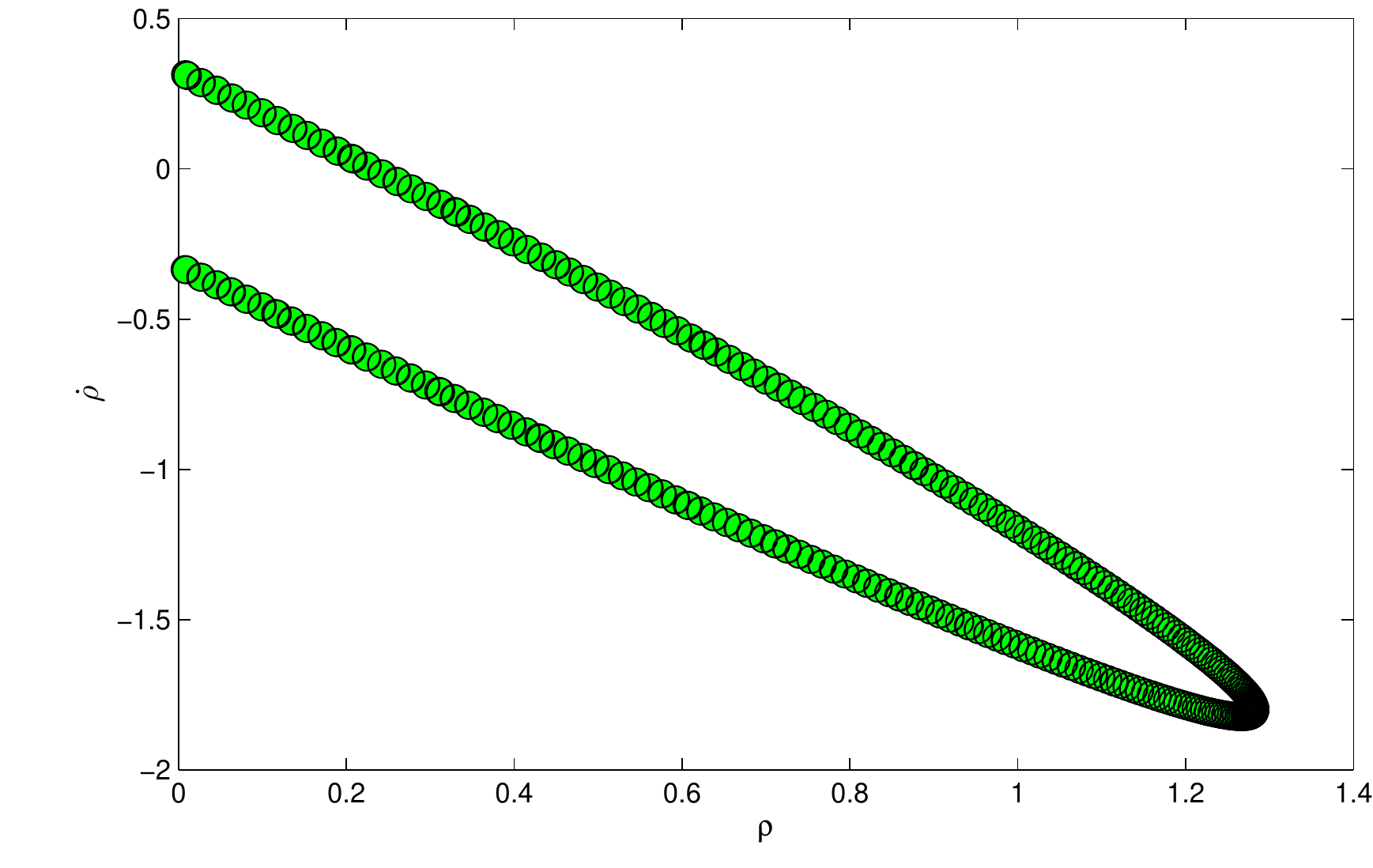}}

\caption{The trajectory we get from the equations of motion for $E=1$ and
$f=0.71$ over the different regions (a) and its Poincare plot using $\omega$ as the period
of the motion (b)}
\end{figure}

We can see that the trajectory (fig.8a) goes into the green areas,
but it doesn't go very deep into it and the Poincare plot (fig.8b)
doesn't suggest chaotic behavior.
Although we can still see a pattern, the Poincare plot spreads on
a much wider range of values now, which indicates that the pattern
is about to break.  

Taking a closer look at the trajectory we see the result (fig.9); 
\begin{figure}
\includegraphics[width=0.8\columnwidth, height=0.8\columnwidth]{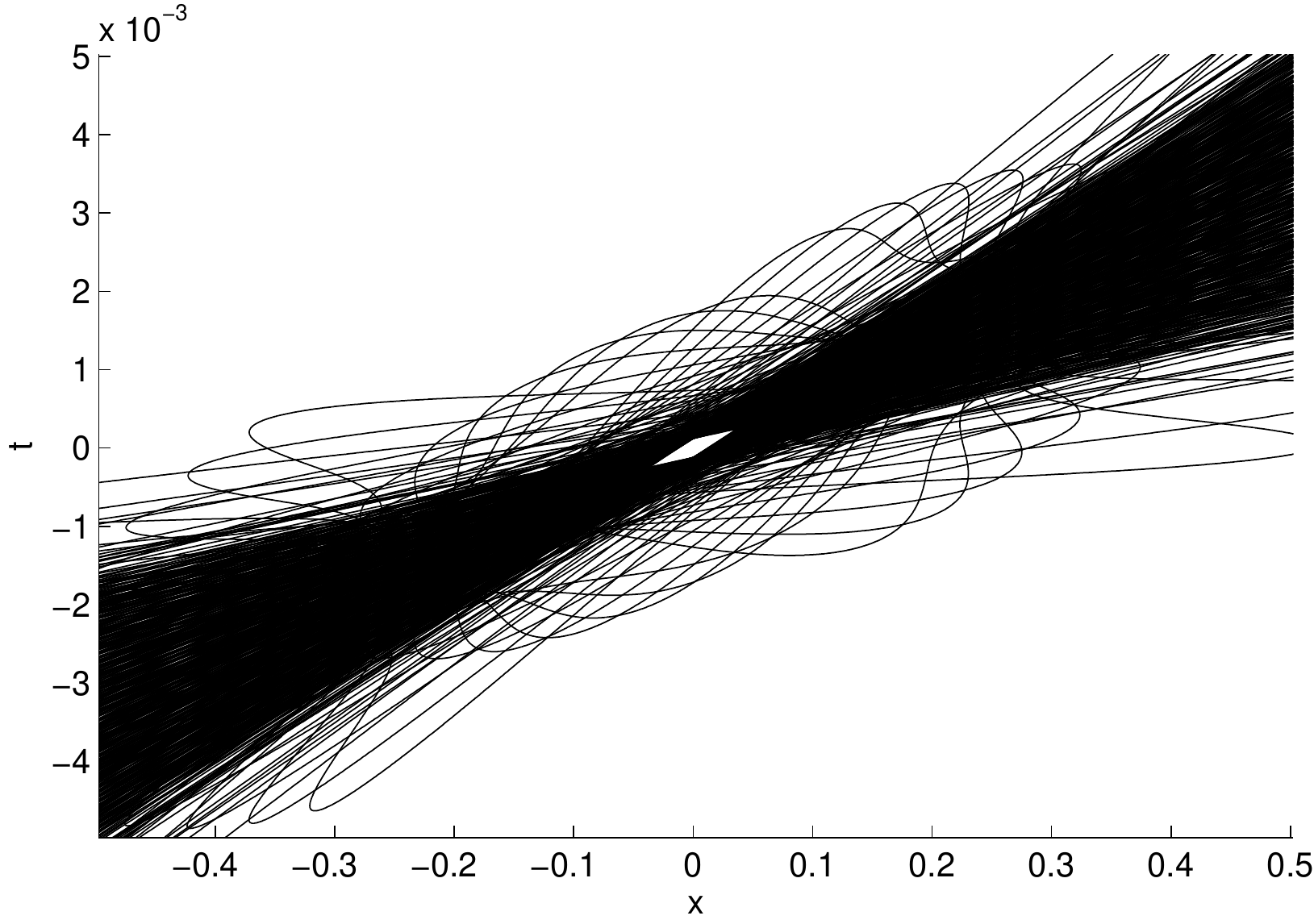}
\caption{A closer look at the trajectory doesn't show the pattern that we identified
for small values of $f$ or the subsets that we saw for large values of $f$}
\end{figure}

This image indicates the beginning of chaotic behavior. We can't see
periodic doubling doubling at this level, but the pattern that we saw for a small value
of $f$ is already broken. 

Also, the trajectories of $x$ vs $\tau$ (fig.10a) and $t$ vs $\tau$
(fig.10b) seems to imply the same.
\begin{figure}
\subfloat[]{\includegraphics[width=0.8\columnwidth, height=0.8\columnwidth]{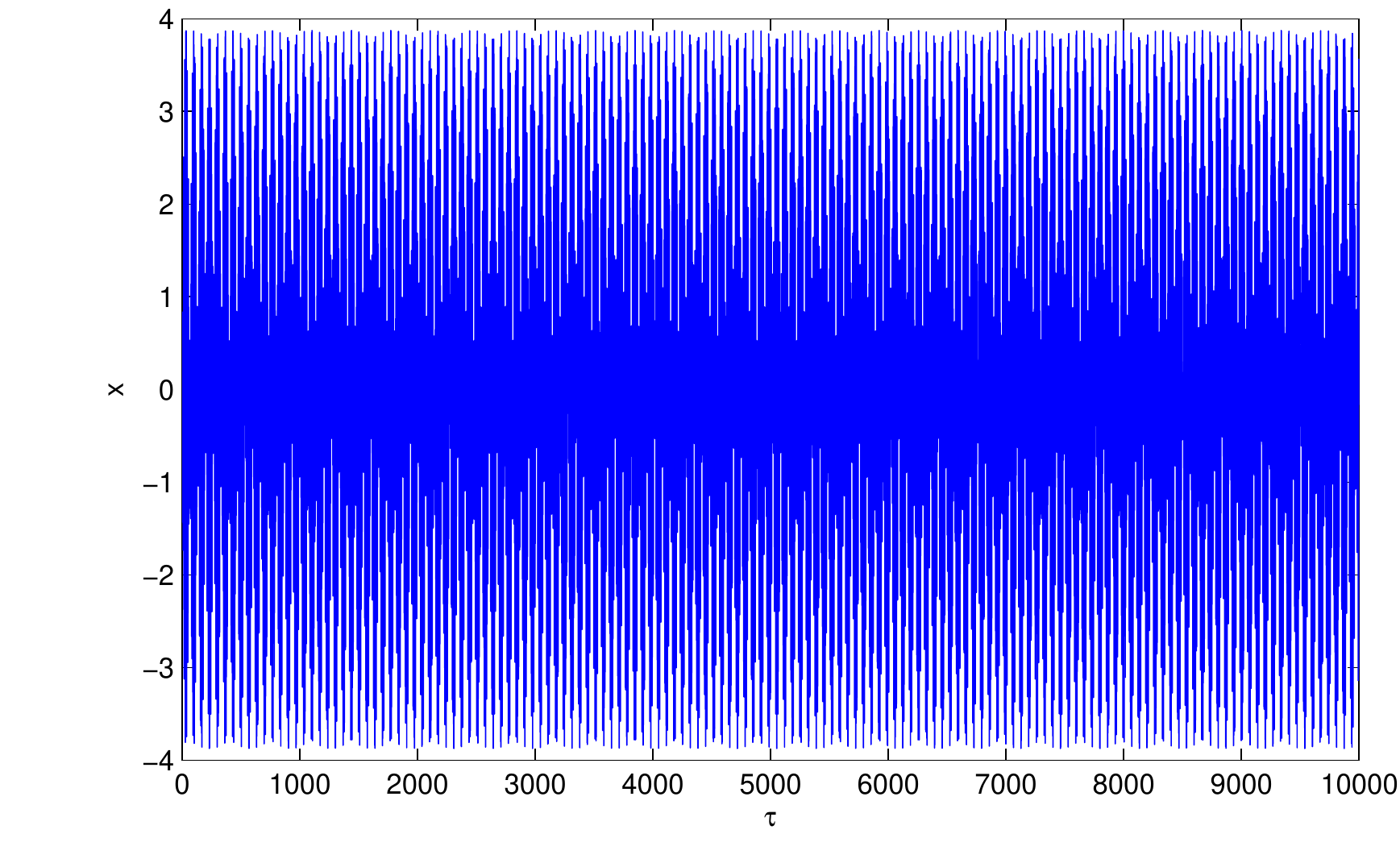}

}\hfill{}\subfloat[]{\includegraphics[width=0.8\columnwidth, height=0.8\columnwidth]{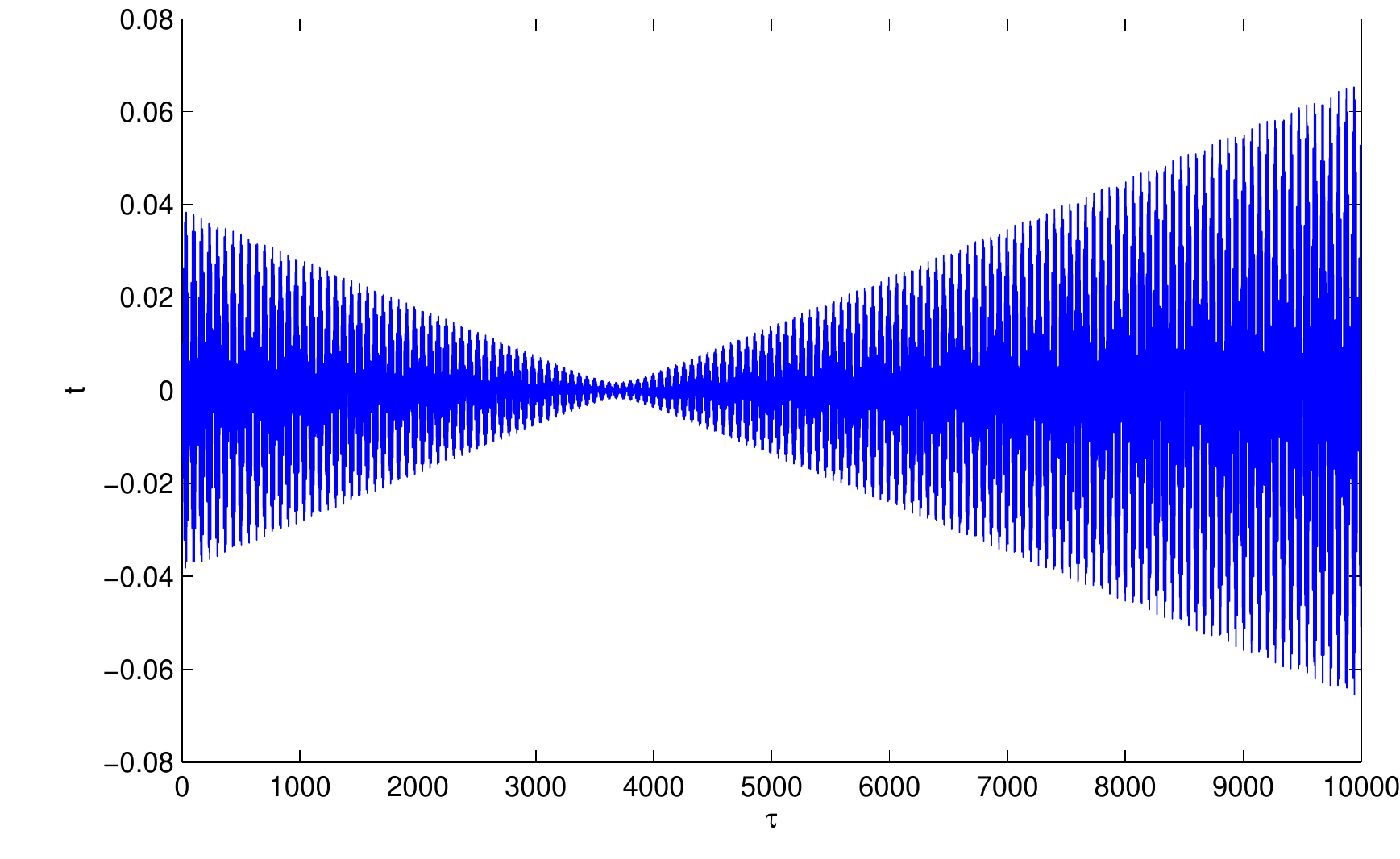}}
\caption{Both $x$ (a) and $t$ (b) as functions of $\tau$ show some pattern and a period in $\tau$. Both the pattern and the period don't seem to be very distinct and it appears that they are about to break.}
\end{figure}

The trajectories don't seem to be chaotic yet, but it looks as if
the pattern is reaching a threshold for chaotic behavior as realized in (fig.5).

\section{Conclusions }

We have studied a method of determining criteria for stablilty of
a relativistic system. First, we performed a conformal transformation
on the original relativistic Hamiltonian which was in flat-space coordinates,
and embedded it into a curved space. Then, we used the
geodesic equation of the curved space in order to obtain a connection
form for the seconed derivative in the original flat space and from
that, we derived a geodesic deviation eqation that depends only on
the flat space coordinates and their derivatives.

\par Next, we used the criteria that we found and applied it to a two
body Hamiltonian in $1+1$ dimensions with a Duffing oscillator 
relative potential $V(y)$,
where $y^\mu = y_1^\mu - y_2^\mu$, with $y_1^\mu $ and $y_2^\mu$ the
coordinates of each of the particles, with a driving force. Our results
describe the {\it relative} motion of the two body system. Since we
are working in $1 +1$ dimensions the corresponding (spacelike) orbital 
motion is
completely described dynamically by the invariant $\rho = \sqrt{y^\mu y_\mu}$
and a hyperbolic angle $\beta$ [14] for which the space component is $y= \rho
\cosh\beta$ and the time component is $t= \rho\sinh\beta$; our results
are given in terms of
the variable  $x$ defined by Eqs. $(14)$ and $(15)$. 
\par We found unstable
regions in the relativitic flat-space and calculated the trajectory
for given initial conditions and a different driving forces numerically.
\par We found two conditions for instability and discovered that there
are two different types of instability: the first is when only one
of the conditions is met and the second is when both conditions are
met. The second type of instability is stronger and if the driving
force is strong enough so that the trajectory enters such a region,
then chaotic behavior appears to be generated. 

We found the threshold driving force value to be $f\approx0.71$.
For this value of $f$, the trajectory enters the region where both
conditions of instability are met and chaotic behavior is starting
to take place. 

\clearpage

\bibliographystyle{unsrt}
\bibliography{art}

\end{document}